\documentstyle[aps,multicol,epsf,epsfig,aps,rotate]{revtex}

\begin{document}

%\draft

\title{Scaling of Cluster and Backbone Mass Between Two Lines in
3d Percolation}

\author{Luciano R. da Silva,$^{\ast\dagger}$ Gerald Paul,$^\ast$
Shlomo Havlin,$^{\ast\ddagger}$ Don R.
Baker,$^{\ast\S}$ and H. Eugene Stanley$^\ast$}

\address{$^\ast$Center for Polymer Studies and Dept. of Physics,
Boston University, Boston, MA 02215 USA\\ 
$\dagger$Departamento de Fisica, UFRN, 59072-970, Natal - RN Brazil\\
$\ddagger$Department of Physics, Bar-Ilan University, Ramat-Gan, Israel\\
$\S$Department of Earth and Planetary Sciences,
McGill University\\ 3450 rue University, Montr\`eal, QC H3A 2A7 Canada}

\date{spbs.tex ~~~ 26 December 2001 ~~~ draft}

\maketitle

\begin{abstract}

We consider the cluster and backbone mass distributions between two
lines of arbitrary orientations and lengths in porous media in three
dimensions, and model the porous media by bond percolation at the
percolation threshold $p_c$. We observe that for many geometrical
configurations the mass probability distribution presents power law
behavior. We determine how the characteristic mass of the distribution
scales with such geometrical parameters as the line length, $w$, the
minimal distance between lines, $r$, and the angle between the lines,
$\theta$. The fractal dimensions of both the cluster and backbone mass
are independent of $w$, $r$, and $\theta$. The slope of the power law
regime of the cluster mass is unaffected by changes in these three
variables. However, the slope of the power law regime of the backbone
mass distribution is dependent upon $\theta$. The characteristic mass of
the cluster also depends upon $\theta$, but the characteristic backbone
mass is only weakly affected by $\theta$. We propose new scaling
functions that reproduce the $\theta$ dependence of the characteristic
mass found in the simulations.

\end{abstract}

\begin{multicols}{2}

\section{Introduction}

\begin{figure}

\centerline{
\epsfxsize=7.0cm
%\epsfclipon
\epsfbox{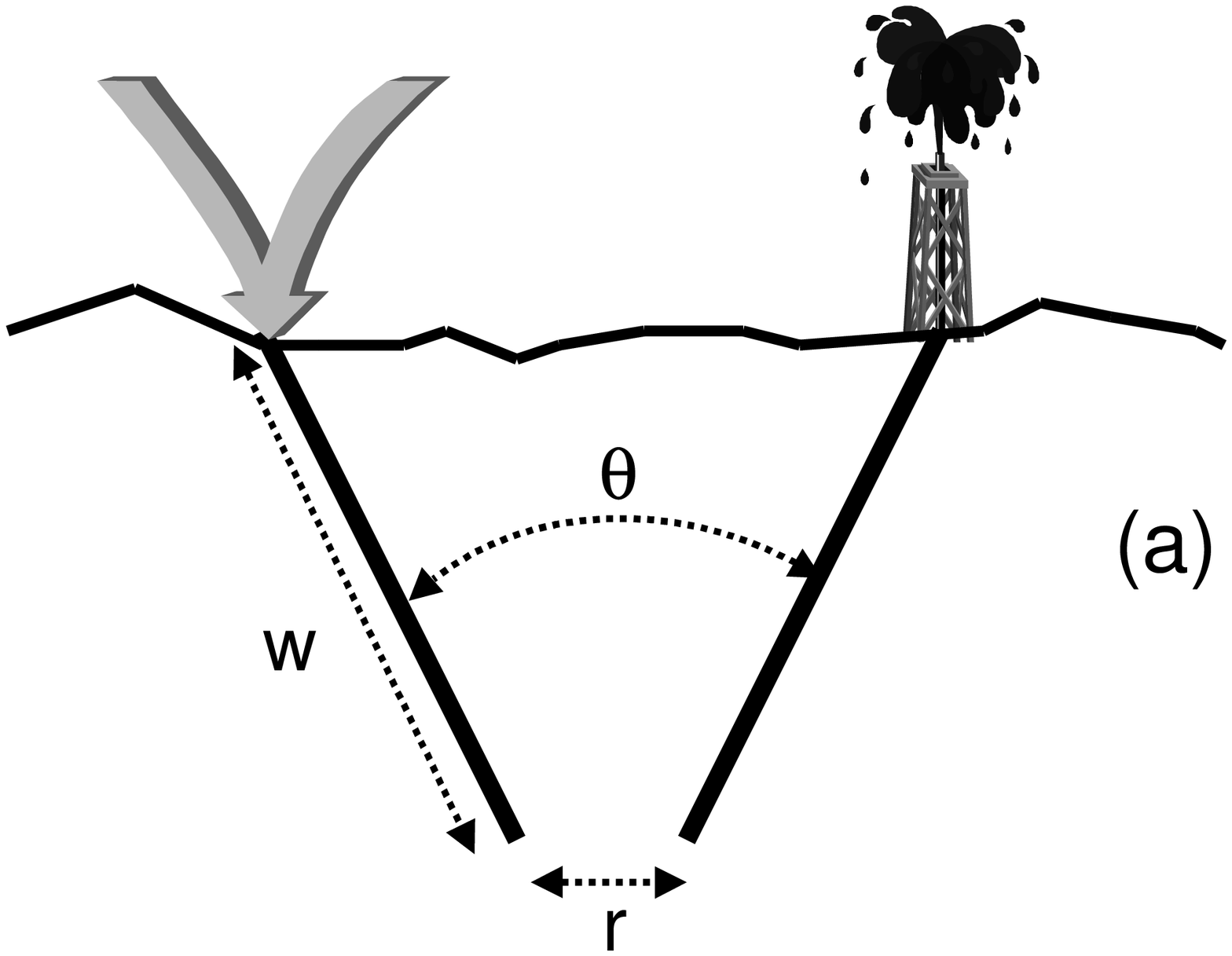}
}

\centerline{
\epsfxsize=8.0cm
%\epsfclipon
\epsfbox{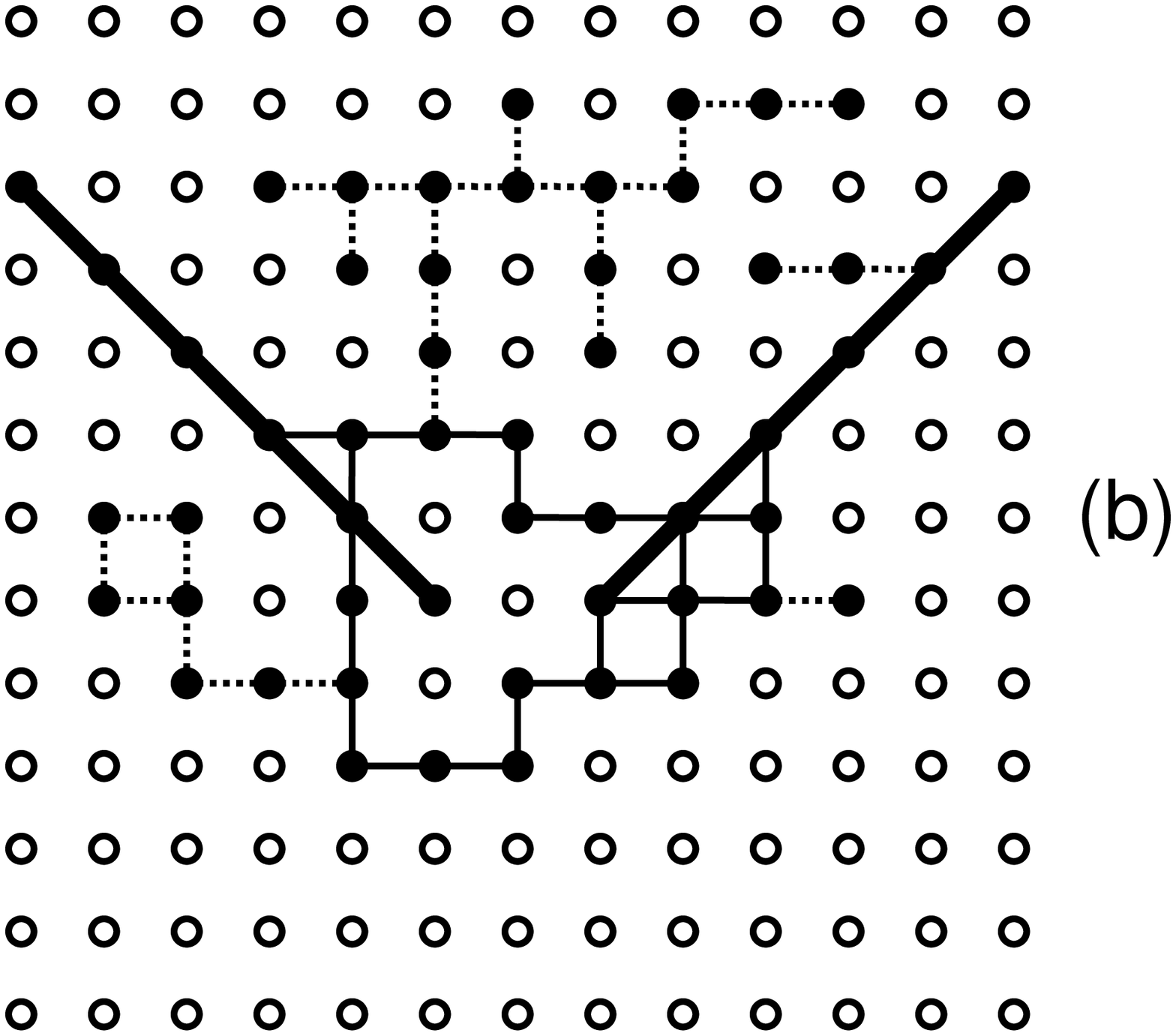}
}

\caption{(a) Illustration of well geometry. (b) Examples of a percolation
cluster with two line wells with parameters $r=2$, $\theta=90^\circ$,
and $w=\sqrt{50}$. The filled sites are members of the percolation cluster, which
has a mass of 52. Solid lines form the backbone, which has a mass of
28.}
\label{examples}
\end{figure}

Since the 1950s, the percolation model has been applied to many
disordered systems \cite{Bunde96,Ben-Avraham00,Stauffer92,Sahimi92}, and
continues to be useful today. Here we use percolation theory to analyze
the cluster and the backbone mass distributions of clusters that are
connected in configurations of the type shown in Fig.~\ref{examples},
configurations in which the two lines are connected by occupied
bonds. The {\it cluster mass\/} is the number of bonds that are
connected to the two lines (Fig.~\ref{examples}a). The {\it backbone\/}
of the cluster is the set of bonds that are connected to the two lines
through independent paths (i.e., paths that have no common bond
\cite{Herrmann84,Stanley77,Coniglio81,Havlin87}). For
configurations of two points, the distributions of various quantities
have been studied
\cite{Dokholyan98,Lee99,Andrade00,Barthelemy99,PaulXX,GrassbergerX,Ziff99X}.
Recently the distribution of the shortest paths between two lines has
been studied for a three-dimensional cubic lattice \cite{PaulXX}, and
here we calculate the mass and backbone distributions.

The motivation for this study is its relevance to techniques of oil
recovery in oil fields \cite{King90}. A common technique used in oil
recovery is the injection of fluid into the ground at one site in the
field in order to force oil out of the ground at another site nearby
(Fig.~\ref{examples}). It is common to inject the fluid along a portion
of the length of the injection well and to collect the oil along a
portion of the length of the production well (as opposed to injecting
and collecting at single points on the wells). In our model, each line
represents a well in the oil field. One line represents the injection
well, and the other the production well. In many cases the oil reservoir
is extremely heterogeneous, and the percolation model is
appropriate. Separation of the rocks into two types---high permeability
(``good rock'') and low or zero permeability (``bad rock'')---can be
accomplished at the outset, with the good rock represented by occupied
bonds and the bad rock represented by unoccupied bonds. The connected
mass represents the total oil in the reservoir connected to the two
wells, and the backbone mass the recoverable oil.

\section{Simulations}

We perform a numerical study of the system using Monte Carlo
simulations. We specify two sets of points representing lines in a
simple cubic lattice to be the wells and we grow the cluster from these
two lines of seeds. If the growth of either cluster stops before the two
clusters connect, we discard the realization. For realizations in which
the two clusters connect, the simulation ends either when the cluster
growth stops naturally, or when the cluster mass reaches some specified
limit. To eliminate finite size effects, we use the techniques of
Ref.~\cite{PaulXX} to simulate systems on lattices of large enough size
that the clusters never reach the edge of the lattice. We perform the
simulations at the percolation threshold, $p_c=0.2488126$
\cite{ZiffXX}. The configurations are characterized by three parameters:
length $w$, angle $\theta$, and minimal distance $r$ [see
Fig.~\ref{examples}(a)]. For each configuration, we run at least $10^6$
non-discarded realizations. We calculate the cluster and backbone mass
for each of these realizations as exemplified in Fig.~\ref{examples}(b).

\section{General Observations}

In both the cluster mass and backbone mass distributions we expect to
observe an initial cutoff due to the fact that these masses cannot be
smaller than the distance $r$. In the backbone distributions, we expect
to see a second cutoff due to the fact that the backbone mass cannot be
greater than the cluster mass at which we stop the simulations. In
addition, for both types of distributions we expect to observe a regime
that exhibits power-law behavior. These general features of the
distributions have been observed in the distributions for other
quantities \cite{Dokholyan98,Lee99,Andrade00,Barthelemy99,PaulXX}.  The
quantities of interest are (i) the most-probable value of the
distribution (the maximum), the scaling of which will be determined by
the fractal dimensions of the quantities measured, and (ii) the slope of
the power-law regime. For clusters grown from a single point, the slopes
of the power law regimes of the cluster and backbone distributions are
$\tau-1$ and $\tau_B-1$, where $\tau$ is the Fisher exponent and
$\tau_B$ the corresponding exponent for the backbone. The fractal
dimensions and power law regime slopes are related by
\cite{Stauffer92,Bunde96,Herrmann84}
\begin{eqnarray}
\label{e1x}
\tau-1 &=& {d\over d_f} \\
\tau_B-1 &=& {d\over d_B},
\end{eqnarray}
where $d$ is the dimension of the system, and $d_f$ and $d_B$ are the
fractal dimensions of the cluster and the backbone, respectively. For
$d=3$, estimates for these exponents are \cite{Strenski91,Rintoul94}
\begin{mathletters}
\begin{eqnarray}
\label{e2xa}
d_f &=& 2.524\pm 0.008 \\
\label{e2xb}
d_B &=& 1.855\pm 0.015 \\
\label{e2xc}
\tau-1 &=& 1.189\pm 0.004 \\
\label{e2xd}
\tau_B-1 &=& 1.617\pm 0.013.
\end{eqnarray}
\end{mathletters}

\section{Cluster Mass}

\subsection{Parallel Wells}
	
\begin{figure}

\centerline{
\epsfxsize=14.0cm
%\epsfclipon
\epsfbox{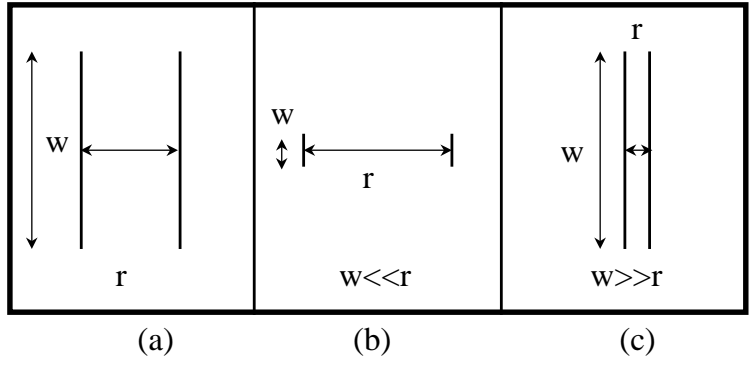}
}

\caption{Parallel well examples. (a) general case (b) $w\ll r$ (c) $w\gg
r$.}
\label{config}
\end{figure}

\begin{figure}

\centerline{
\epsfxsize=7.0cm
\epsfclipon
\rotate[r]{\epsfbox{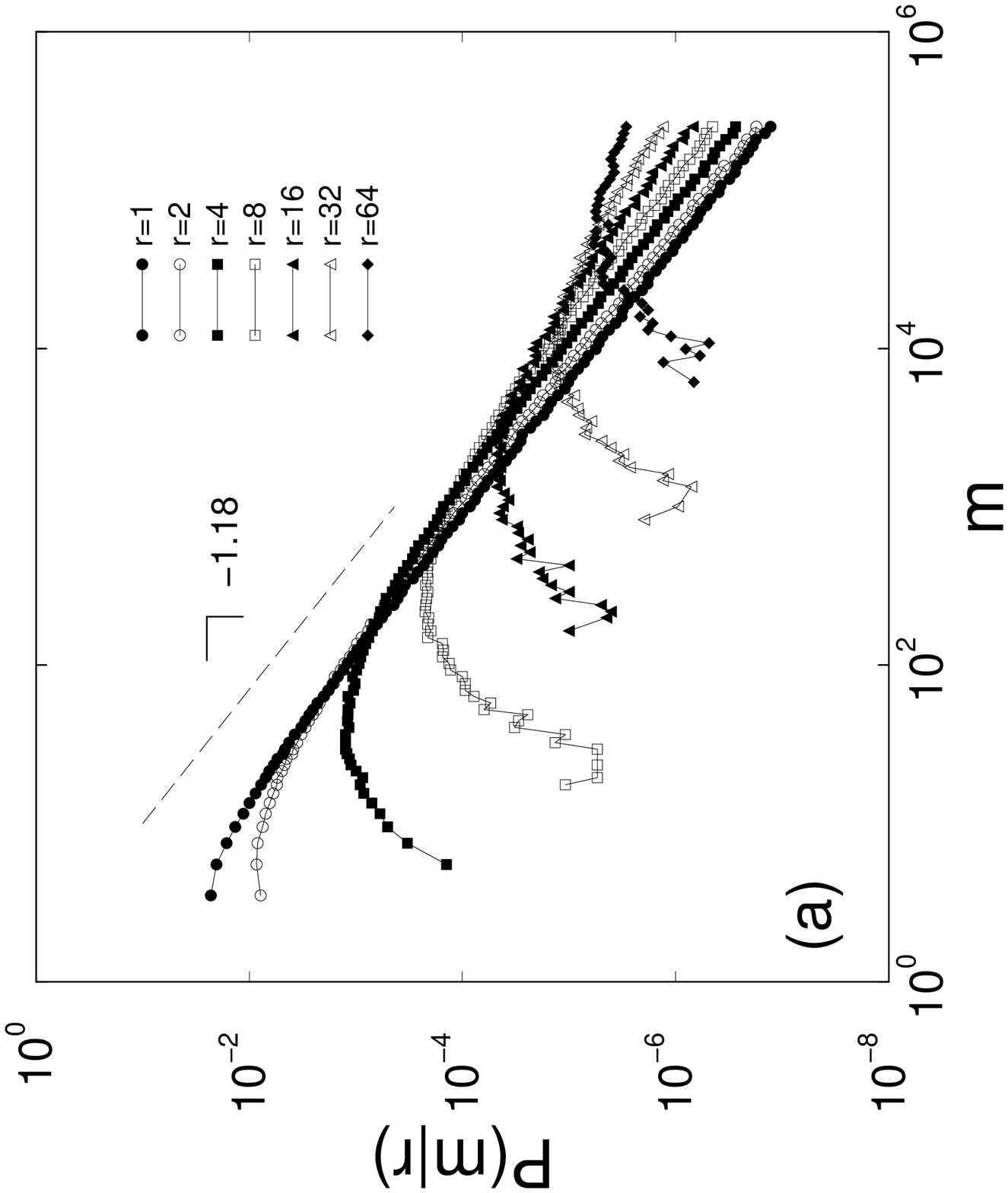}}
}

\centerline{
\epsfxsize=7.0cm
\epsfclipon
\rotate[r]{\epsfbox{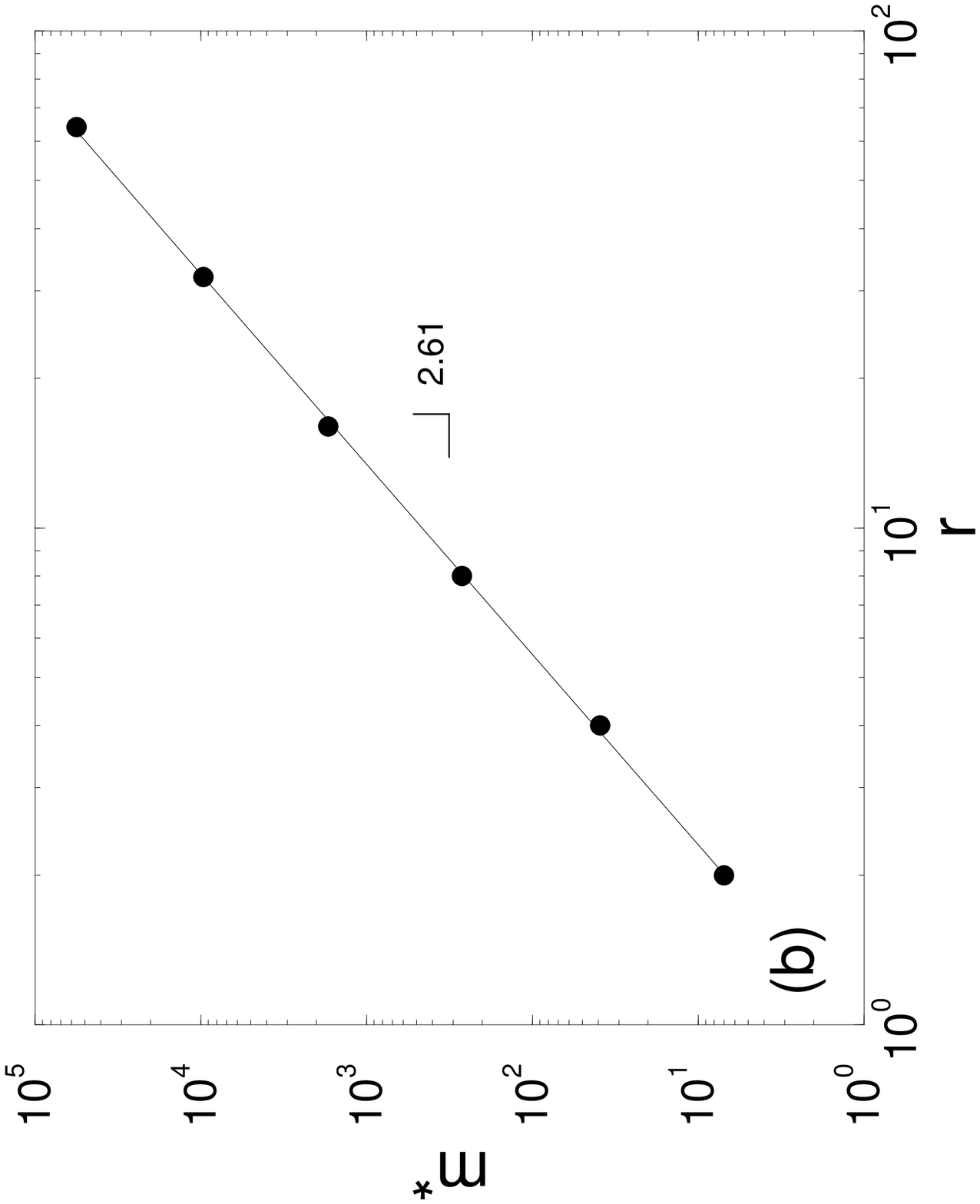}}
}

\caption{(a) Cluster mass distribution $P(m|r)$, for the percolation
cluster for the two points case ($w=0$), for several values of $r$
($r=1,2,4,8,16,32,64$). The line of slope $-1.18$ denotes the theoretical
expectation. (b) Scaling behavior of $m^\ast$, the most probable mass,
as a function of $r$. The fitted slope of 2.61 is consistent with the
fractal dimension $d_f=2.54$}
\label{massrXw0}
\end{figure}

\begin{figure}

\centerline{
\epsfxsize=7.0cm
\epsfclipon
\rotate[r]{\epsfbox{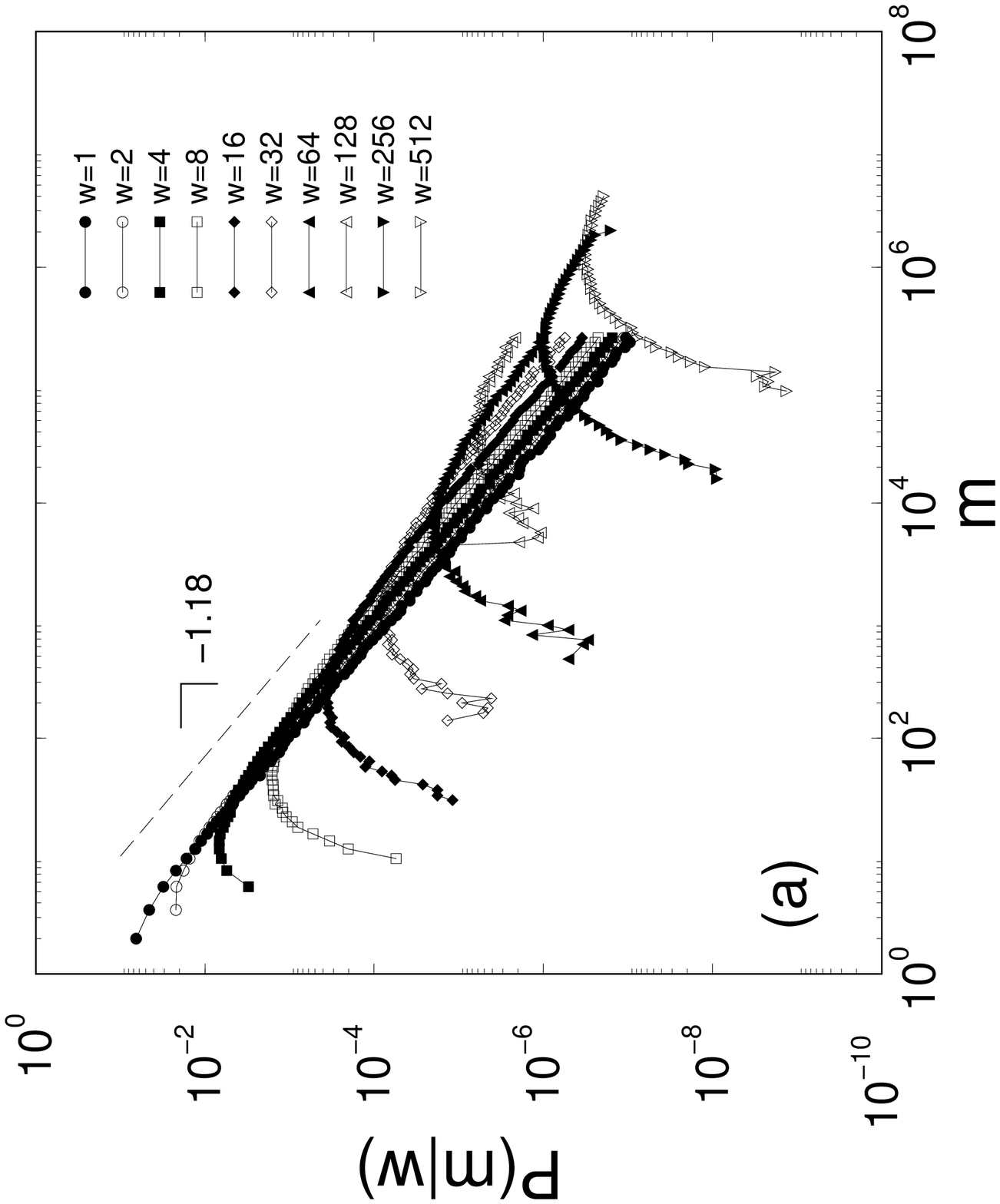}}
}

\centerline{
\epsfxsize=7.0cm
\epsfclipon
\rotate[r]{\epsfbox{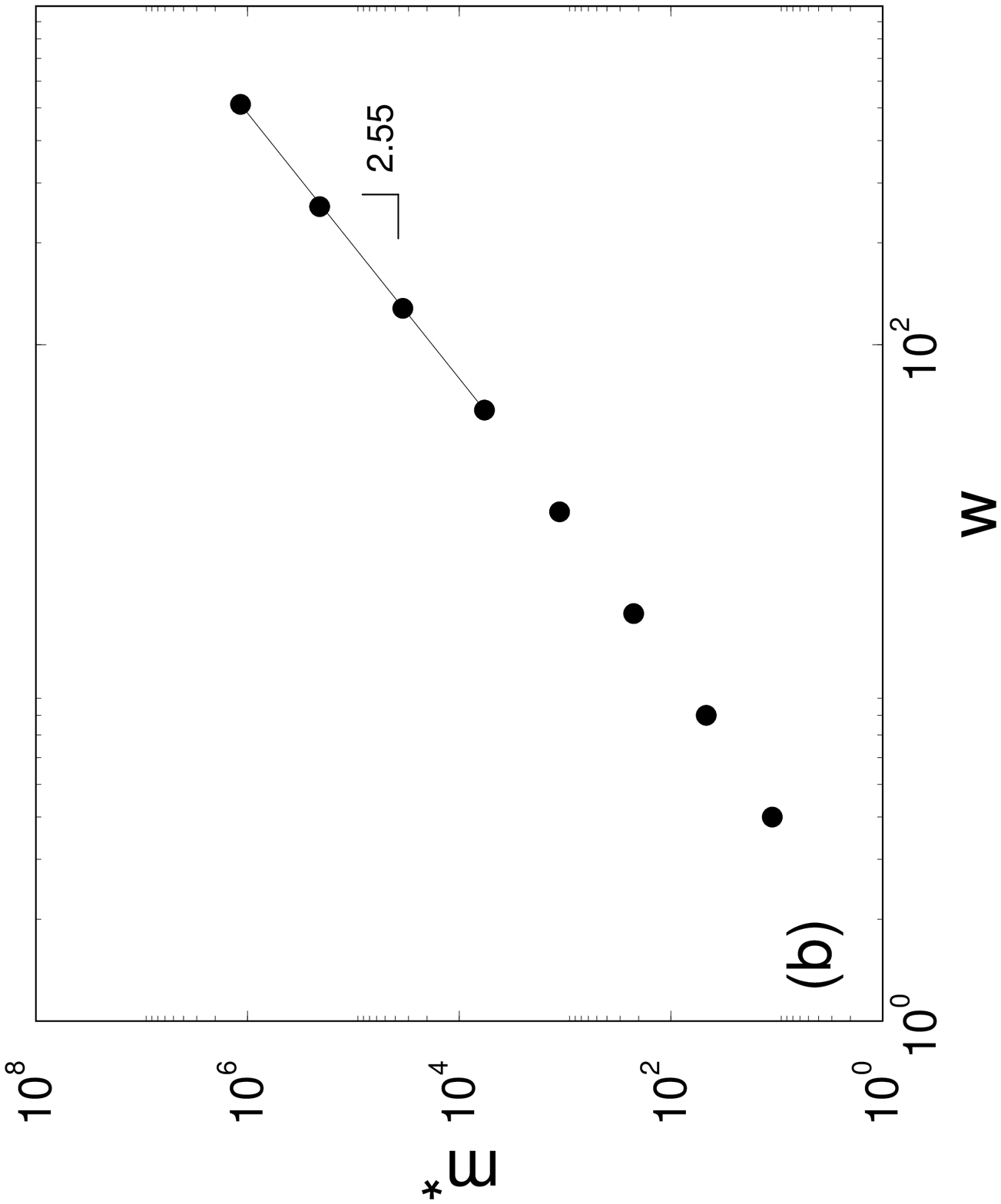}}
}

\caption{(a) Cluster mass distribution $P(m|w)$ for ($r=0$) for several
values of $w$ ($w=1,2,4,8,16,32,64,128,256,512$). (b) Scaling behavior
of $m^\ast$ as a function of $w$.}
\label{massr0WX}
\end{figure}

In order to gain insight into the general behavior, we first study
parallel wells ($\theta=0$) see Fig.~\ref{config}(a). We consider first
the following limiting cases:

\begin{itemize}

\item[{(i)}] $w\ll r$ --- In this case we approximate the configuration
by two points (see Fig.~\ref{config}(b)). In Fig.~\ref{massrXw0}(a) we
show the mass probability distribution $P(m|r)$ for $w=0$ and $r=1$, 2,
4, 8, 16, 32 and 64. The distribution shows a maximum followed by a
power-law regime with slope $-1.18$, consistent with Eq.~(\ref{e2xc}).
We study also how the characteristic mass $m^\ast$, corresponding to
peak of the distribution, scales with the distance $r$. The log-log plot
of $m^\ast$ vs $r$ in Fig.~3b indicates that $m^\ast$ scales with
exponent $d_B\cong 2.6$ which is consistent with Eq.~(\ref{e2xa}).
	
\item[{(ii)}] $w\gg r$ --- For this case [see Fig.~\ref{config}(c)] we
approximate the configurations by $r=0$ (a single line). We perform
the same analysis as before, and obtain similar results [see
Fig.~\ref{massr0WX}(a)], i.e., power law distribution for $P(m|w)$ with
a slope $\approx -1.18$ and fractal dimension $\approx 2.55$
[Fig.~\ref{massr0WX}(b)].

\end{itemize}

We now study cases intermediate to those studied in (i) and (ii). In
Fig.~5a, we plot the distribution of cluster mass for configurations in
which $r=16$ and we vary $w$ from 0 to 64. For small $w$, the
distributions are essentially unchanged, but for $w\gg r$, the
distributions scale with the exponent $d_f$ (Fig.~5a and 5b).

We now develop a scaling form for the dependence of the characteristic
mass $m^\ast$ on $r$ and $L$, the system size. Without loss of
generality we can write
\begin{equation}
m^\ast(r,L)=\left[f\left({r\over w}\right)r\right]^{d_f}.
\label{e1}
\end{equation}
This form is consistent with the scaling of the cluster mass. That is,
if
\begin{eqnarray}
\nonumber r' &\equiv& \alpha r \\
     w' &\equiv& \alpha w, 
\label{e2}
\end{eqnarray}
then
\begin{equation}
m^\ast(r',L')=\left[f\left({r'\over w'}\right)r'\right]^{d_f}=\left[f\left({r\over
w}\right)\alpha r\right]^{d_f}=\alpha^{d_f}m^\ast(r,L).
\label{e3}
\end{equation}
We further assume that $m^\ast(r,L)$ can be written as 
\begin{equation}
m^\ast(r,L)=\left[ar+g\left({r\over w}\right)w\right]^{d_f},
\label{e4}
\end{equation}
since $w$ and $r$ become irrelevant variables for $r\gg w$ and $r\ll w$,
respectively. Thus we expect
\begin{equation}
g\left({r\over w}\right)\to\cases{
0        & $r\gg w$\cr
\mbox{constant} & $r\ll w$}.
\label{e5}
\end{equation}

\subsection{Non-Parallel Wells}

We now study non-parallel wells. The results for the mass probability
distribution $P(m|\theta)$ are shown in Fig.~\ref{massr8thetaX}(a). We
find that the power law regime is consistent with a slope $-1.18$
independent of $\theta$. We consider also the peak of these
distributions, analyzing how $m^\ast$ evolves with $\theta$. Figure
\ref{massr8thetaX}(b) shows the dependence of $m^\ast$ versus
$\theta$. In this case we do not find a power distribution for the
scaling behavior $m^\ast$ versus $\theta$. $m^\ast$ increases rapidly
for small values of $\theta$, and for larger $\theta$ asymptotically
approaches a limiting value at $\theta=\pi$.

We now suggest a functional form for the dependence of the
characteristic mass $m^\ast$ on $\theta$. Without loss of generality we
can write
\begin{equation}
m^\ast(\theta)=m^\ast(0)[f(\theta)]^{d_f}.
\label{e6}
\end{equation}
Since the configuration for $\theta=\pi$ is simply a single straight
line twice the length of the single line for $\theta=0$, we expect
\begin{equation}
m^\ast(\pi)=m^\ast(0)2^{d_f}.
\label{e7}
\end{equation}
We then are motivated to write
\begin{equation}
m^\ast(\theta)=m^\ast(0)[1+g(\theta)]^{d_f},
\label{e8}
\end{equation}
where $g(\theta)$ is monotonic and 
\[
g(0)=0
\]
and
\[
g(\pi)=1.
\]

A first guess at a functional form for $g(\theta)$ is some power of
$\sin(\theta/2)$ but no power seems to fit the data of
Fig.~\ref{massr8thetaX}(b) well. A functional form that fits better is
\begin{equation}
h(\theta)=\sin\left[{\pi\over 2}\sin\left({\theta\over
2}\right)\right]^{0.4}. 
\label{e9}
\end{equation}
The final functional form for $m^\ast(\theta)$ is thus
\begin{equation}
m^\ast(\theta)=m^\ast(0)\left(1+\sin\left[{\pi\over
2}\sin\left({\theta\over 2}\right)\right]^{0.4}\right)^{d_f},
\label{e9x}
\end{equation}
where the exponent 0.4 is obtained by the power law fit in
Fig.~\ref{massthetafit}(a). We note that there is no {\it a priori\/}
justification for this form; it simply satisfies the appropriate
boundary conditions and fits the simulation results reasonably well, as
shown in Fig.~\ref{massthetafit}(b).

\section{Backbone Mass}

We proceed to analyze the scaling of the backbone mass distributions as
we did for the cluster mass.

\subsection{Parallel Wells}

\begin{itemize}

\item[{(i)}] $w\ll r$ --- We present in Fig.~\ref{bbrXw0}(a) the
backbone probability distribution $P(m_B|r)$. As seen in
Fig.~\ref{bbrXw0}(a) and (b), the power law regimes of these plots are
consistent with Eq.~(\ref{e2xd}), and the distributions scale with an
exponent consistent with $d_B$. Similar results have been found
previously \cite{Barthelemy99} but are shown here for completeness.

\item[{(ii)}] $w\gg r$ --- We show in Fig.~\ref{bbr0WX}(a) the backbone
probability distribution $P(m_B|w)$ for this case. Again, the power law
regimes of these plots are consistent with Eq.~(\ref{e2xd}) and the
distributions scale asymptotically with an exponent consistent with
$d_B$ [Fig.~\ref{bbr0WX}(b)] although there are large
corrections-to-scaling as seen in the small $w$ behavior of
Fig.~\ref{bbr0WX}(b).

\end{itemize}

Figures \ref{bbr0WX}(a) and \ref{bbr0WX}(b) are also indicative of the
behavior of cases intermediate to the limiting cases of (i) and (ii)
above. For $w\ll r$, changing $w$ will have essentially no effect on the
distributions; for $w\gg r$, the distributions will scale with $d_B$ as
$w$ is changed.

\subsection{Non-Parallel Wells}

We show the results for the backbone probability distribution
$P(m_B|\theta)$ in Fig.~\ref{bbr8thetaX}(a) for various values of
$\theta$. In contrast to the distributions of cluster mass, the
distributions of backbone mass exhibit power-law regimes, the exponents
of which depend on $\theta$ while the characteristic mass $m^\ast$, is
essentially constant as a function of $\theta$. Thus an appropriate
functional form for $P(m_B|\theta)$ is
\begin{equation}
P(m_B|\theta)\sim\left({m_B\over
r^{d_B}}\right)^{g_B(\theta)}f_1\left({m_B\over
r^{d_B}}\right)f_2\left({m_B\over L^{d_B}}\right),
\label{e10x}
\end{equation}
where $f_1$ and $f_2$ are cutoff functions. The first cutoff function,
$f_1$, reflects the fact that the backbone mass must always be at least
equal to the distance $r$ between the two points; the second cutoff
function, $f_2$, reflects the fact that the backbone mass is bounded
because of the finite size, $L$, of the system. Similar behavior has
been observed for the distributions of shortest paths between two lines
\cite{PaulXX}.

In order to determine the varying slope more accurately we perform
simulations for various $\theta$ for $r=1$, which results in the largest
power-law regime. The results of these simulations are shown in
Fig.~\ref{bbr8thetaX}(b). In Fig.~\ref{bbr8thetaX}(c) we plot the
power-law regime exponent, $g_B$ vs $\theta$. For $\theta=0$, the
configuration is that of parallel lines, and the exponent is
$\tau_B$. For $\theta=\pi$, the exponent decreases to a value of about
0.84. The marked difference in these two exponents is shown clearly in
Fig.~11, in which we plot for fixed $r=1$, $\theta=180^\circ$, $P(m_B|r)$ for
various values of $w$. The larger the value of $w$, the later a
crossover occurs from behavior reflecting a configuration of 2 lines
with $\theta=180^\circ$ to a configuration effectively of 2 points, with
power-law-regime exponent $\tau_B -1$.

\section{Discussion}

We have analyzed the distributions of cluster mass and backbone mass for
various configurations of 2-line 3d percolation clusters. The behavior of the
cluster mass distributions is not remarkable. On the other hand, we have
found that the exponent of the power-law-regime for the backbone mass
distributions is dependent on the angle $\theta$ between the lines. It
remains to develop a theory which can predict the specific dependence of
this exponent on $\theta$.

\subsubsection*{Acknowledgements}

We thank L. Braunstein and S. V. Buldyrev for helpful discussions, and
British Petroleum, CPNq, and the National Science Foundation for support.

\end{multicols}
	
\newpage

\newpage

\newpage

\newpage

\newpage

\begin{figure}

\centerline{
\epsfxsize=7.0cm
\epsfclipon
\rotate[r]{\epsfbox{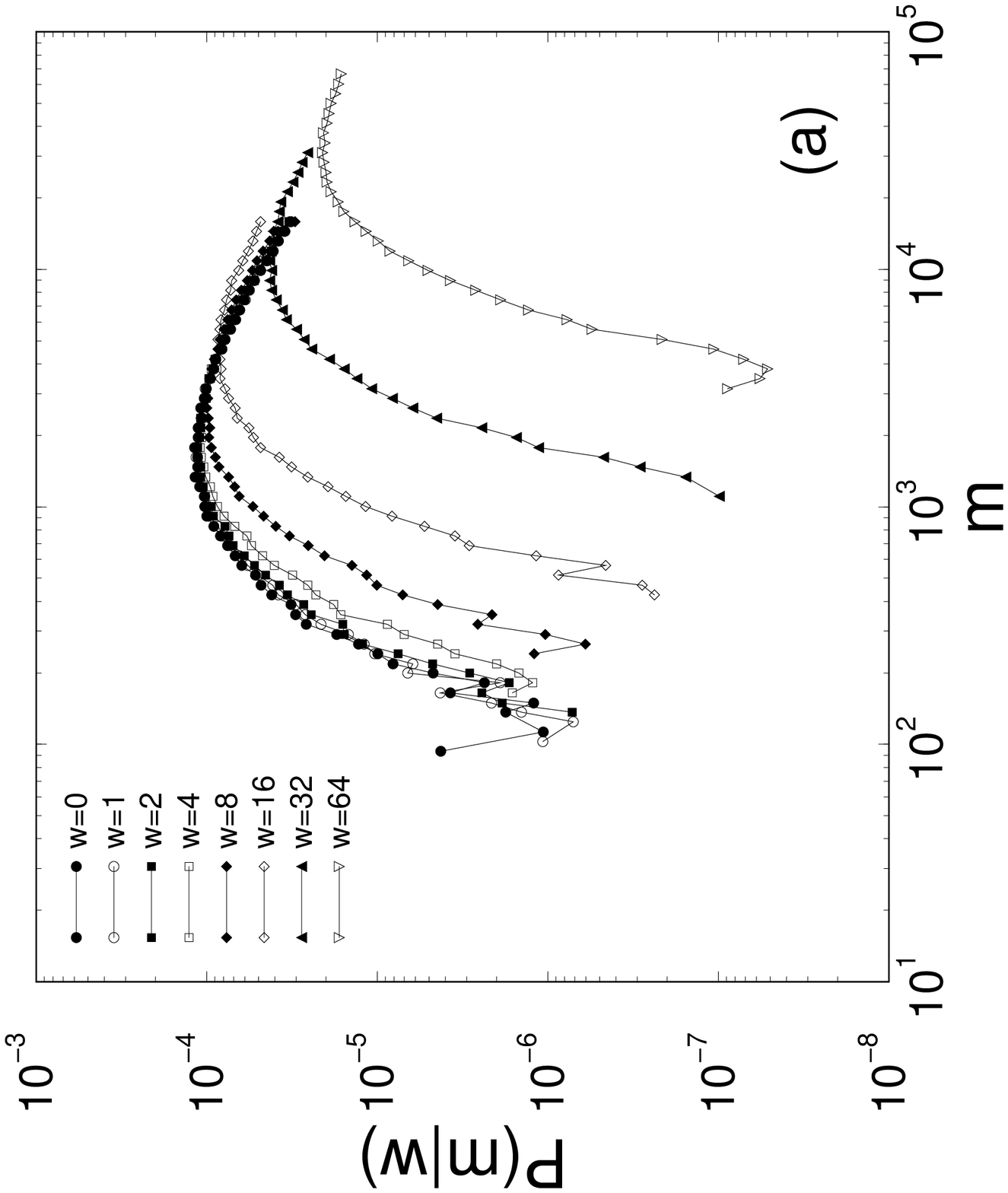}}
}

\centerline{
\epsfxsize=7.0cm
\epsfclipon
\rotate[r]{\epsfbox{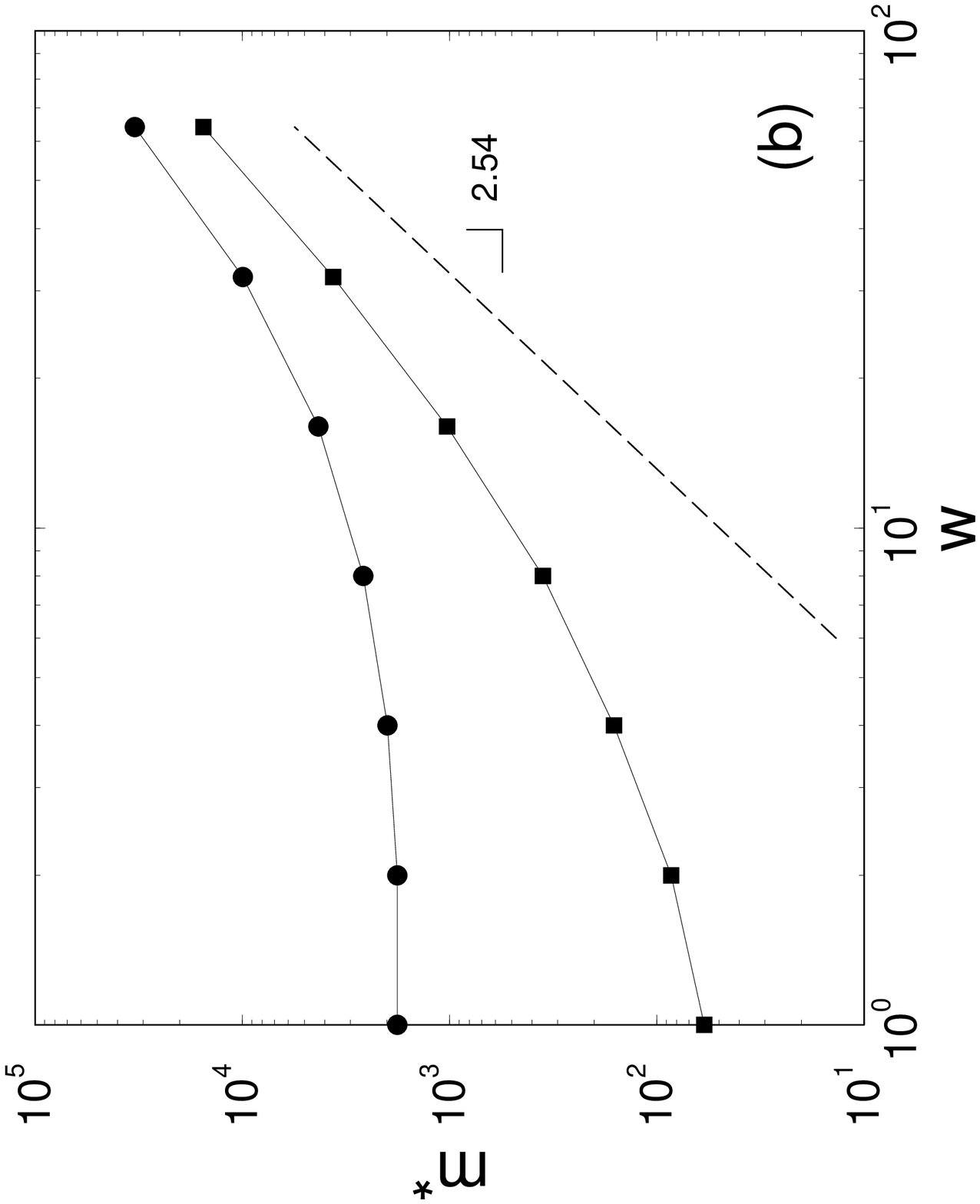}}
}

\caption{(a) Mass distribution $P(m|w)$ for two parallel wells with
$r=16$ and $w=0,1,2,4,8,16,32,64$. (b) Scaling behavior of $m^\ast$
versus $w$ for $r=4$ (circles) and for $r=16$ (squares).}
\label{masscutoffs}
\end{figure}

\newpage

\begin{figure}

\centerline{
\epsfxsize=7.0cm
\epsfclipon
\rotate[r]{\epsfbox{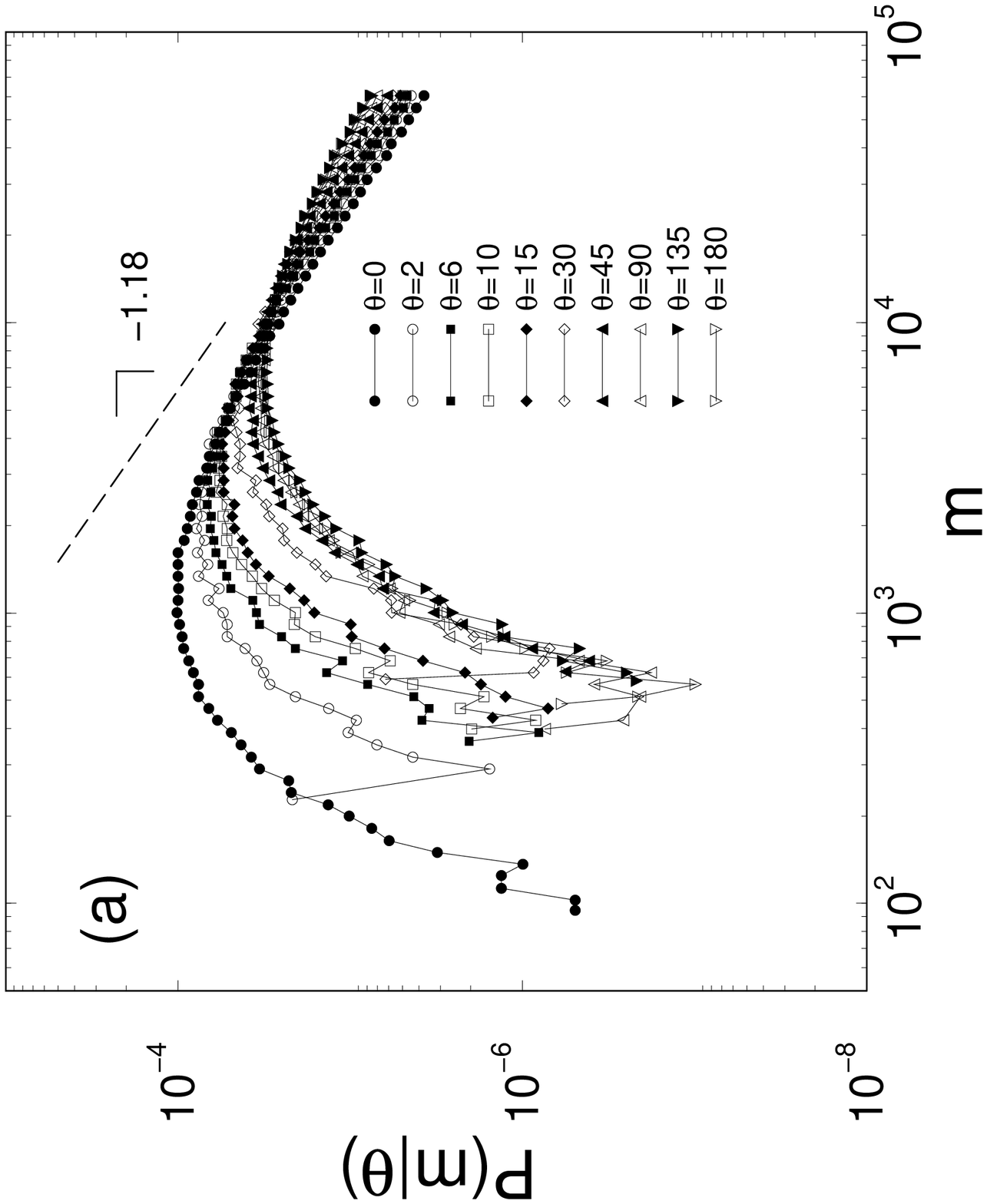}}
}

\centerline{
\epsfxsize=7.0cm
\epsfclipon
\rotate[r]{\epsfbox{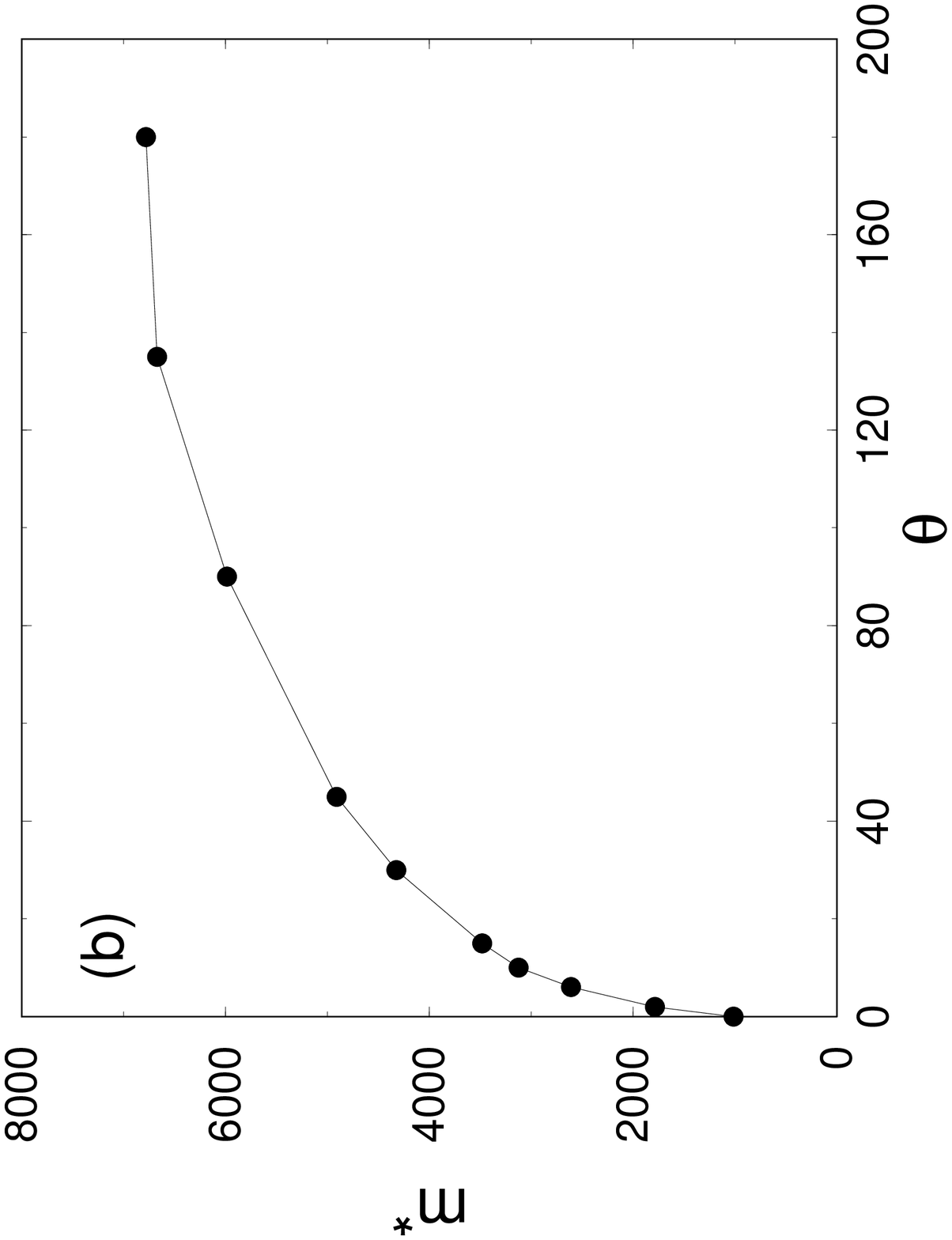}}
}

\caption{(a) Mass distribution $P(m|\theta)$ for $r=0$ and $w=32$ for
the percolation cluster to the general non parallel wells for several
values of the angle $\theta$.  (b) Corresponding scaling behavior of
$m^\ast$ versus $\theta$.}
\label{massr8thetaX}
\end{figure}

\newpage

\begin{figure}

\centerline{
\epsfxsize=7.0cm
\epsfclipon
\epsfbox{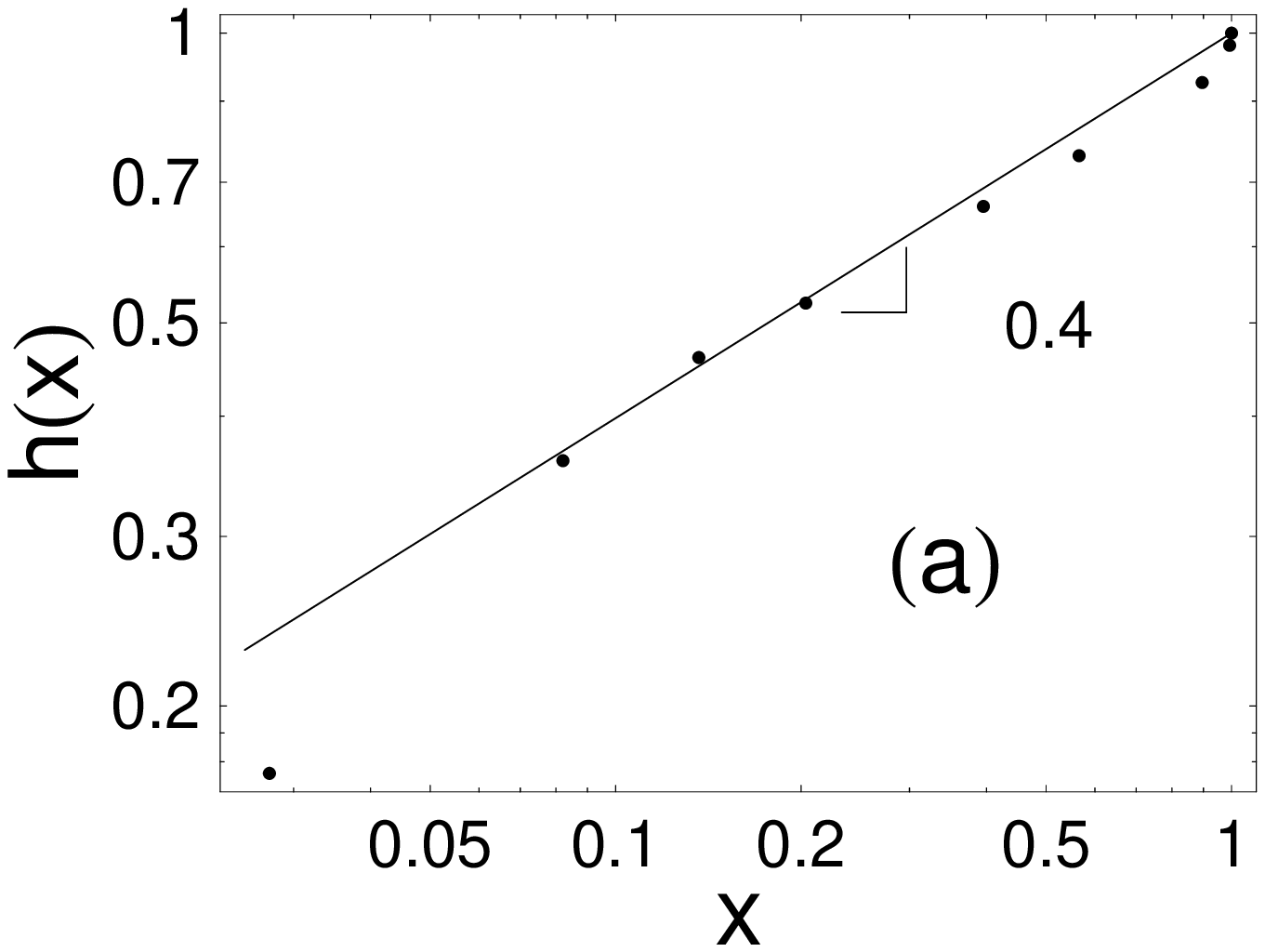}
}

\centerline{
\epsfxsize=7.0cm
\epsfclipon
\epsfbox{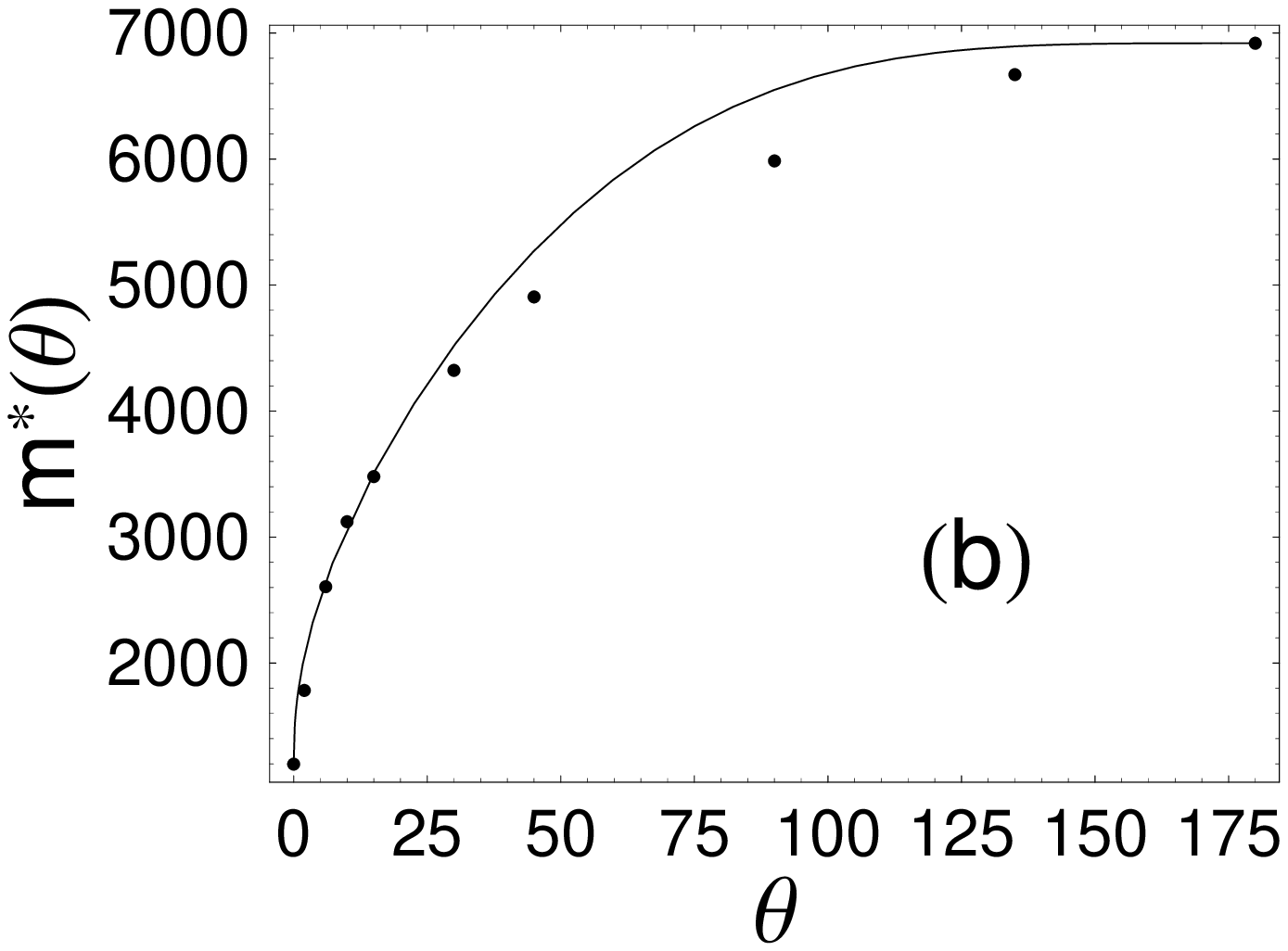}
}

\caption{(a) Determination of exponent in Eq.~(12), where
$x\equiv\sin[(\pi/2)\sin(\theta/2)]$. (b) Comparison of functional form
$m(\theta)$ (solid line) versus observed data.}
\label{massthetafit}
\end{figure}

\newpage

\begin{figure}

\centerline{
\epsfxsize=7.0cm
\epsfclipon
\rotate[r]{\epsfbox{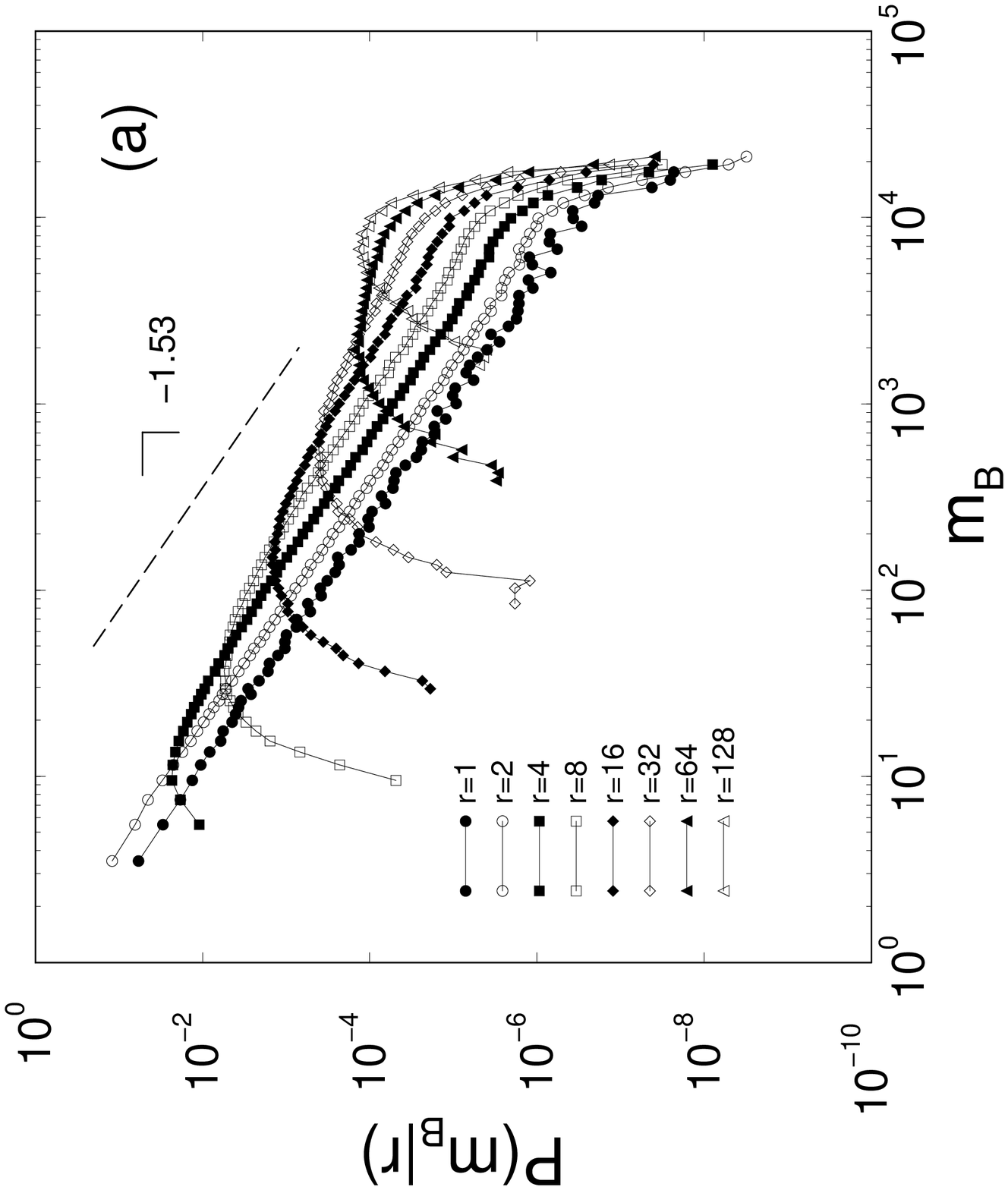}}
}

\centerline{
\epsfxsize=7.0cm
\epsfclipon
\rotate[r]{\epsfbox{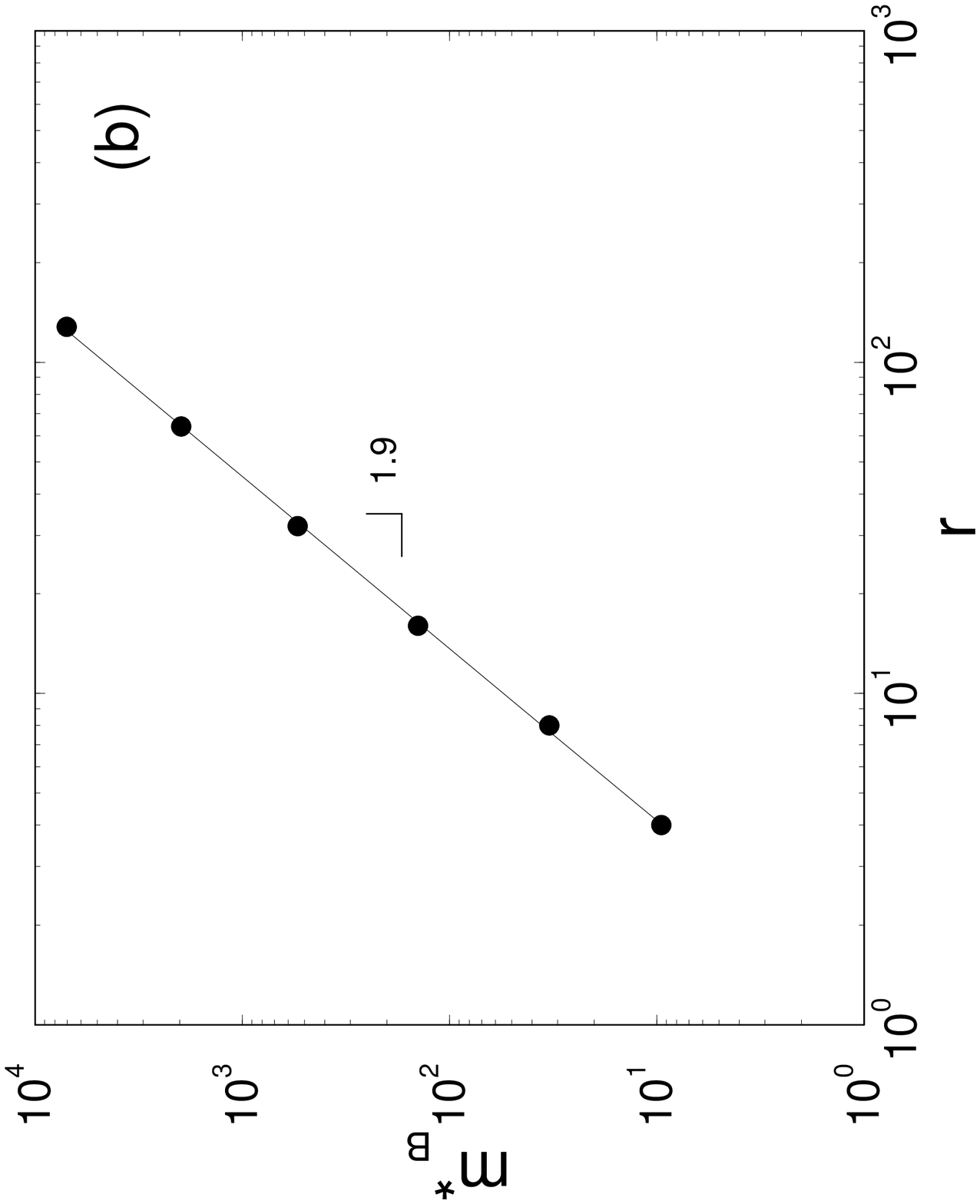}}
}

\caption{(a) Backbone mass distribution $P(m_B|r)$ for the percolation
cluster for two points and $r=1,2,4,8,16,32,64,128$ (b) Scaling behavior
of $m_B^\ast$ versus $r$.}
\label{bbrXw0}
\end{figure}

\newpage

\begin{figure}

\centerline{
\epsfxsize=7.0cm
\epsfclipon
\rotate[r]{\epsfbox{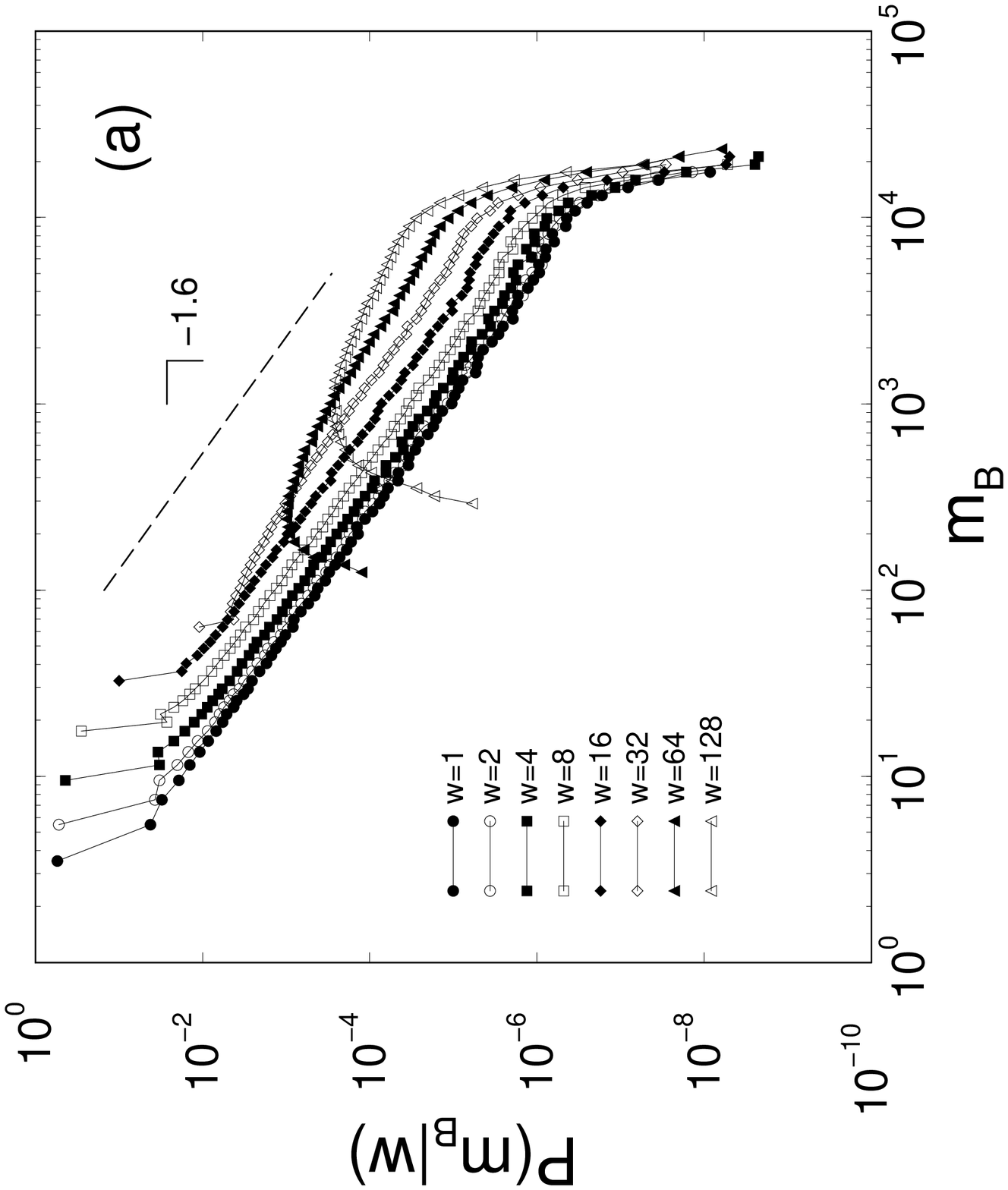}}
}

\centerline{
\epsfxsize=7.0cm
\epsfclipon
\rotate[r]{\epsfbox{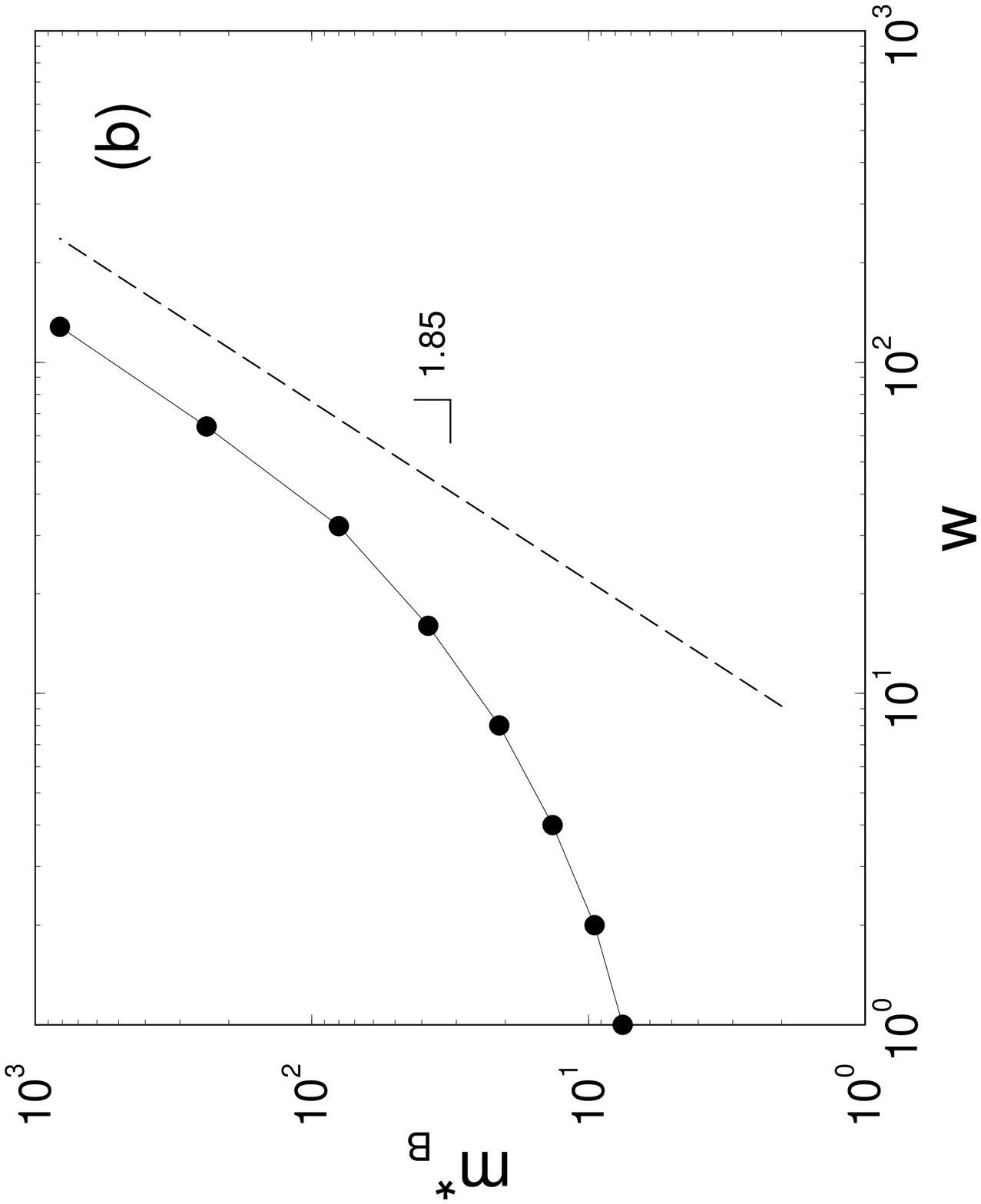}}
}

\caption{(a) Backbone mass distribution $P(m_B|w)$ for parallel wells
for $r=1$ and $w=1,2,4,8,16,32,64,128$. (b) The corresponding scaling behavior
of $m_B^\ast$ versus $w$.}
\label{bbr0WX}
\end{figure}

\newpage

\begin{figure}

\centerline{
\epsfxsize=7.0cm
\epsfclipon
\rotate[r]{\epsfbox{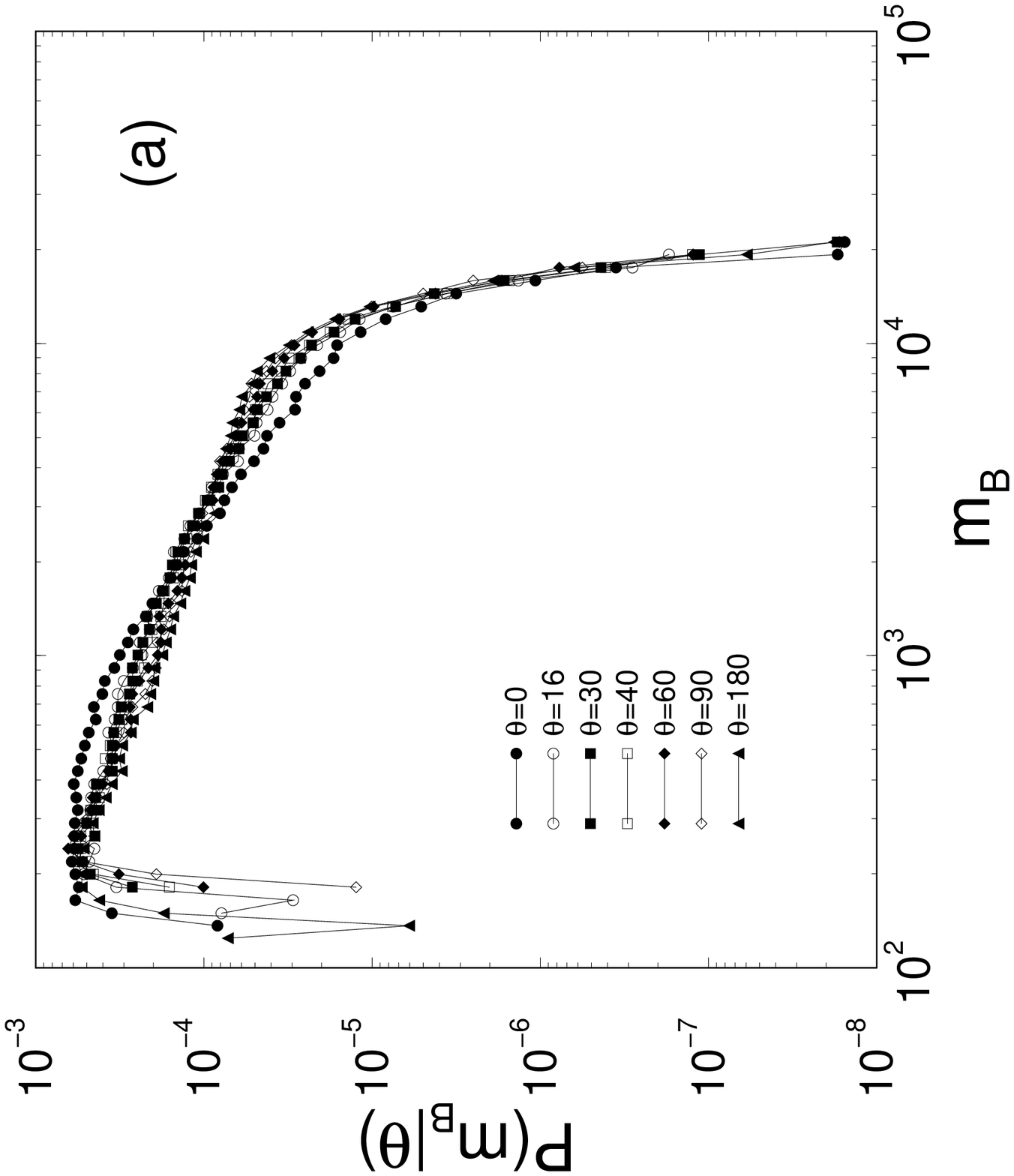}}
}

\centerline{
\epsfxsize=7.0cm
\epsfclipon
\rotate[r]{\epsfbox{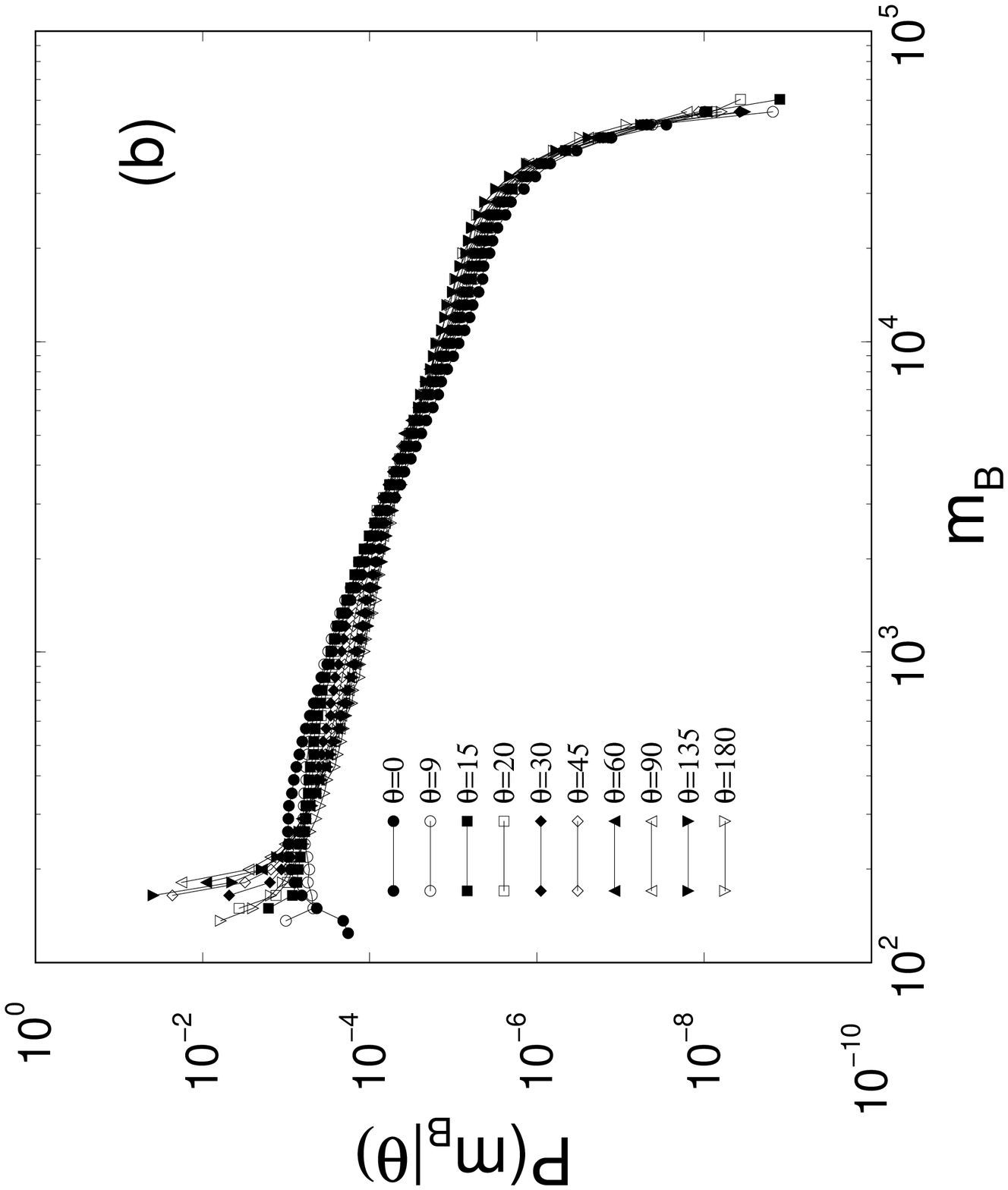}}
}

\centerline{
\epsfxsize=7.0cm
\epsfclipon
\rotate[r]{\epsfbox{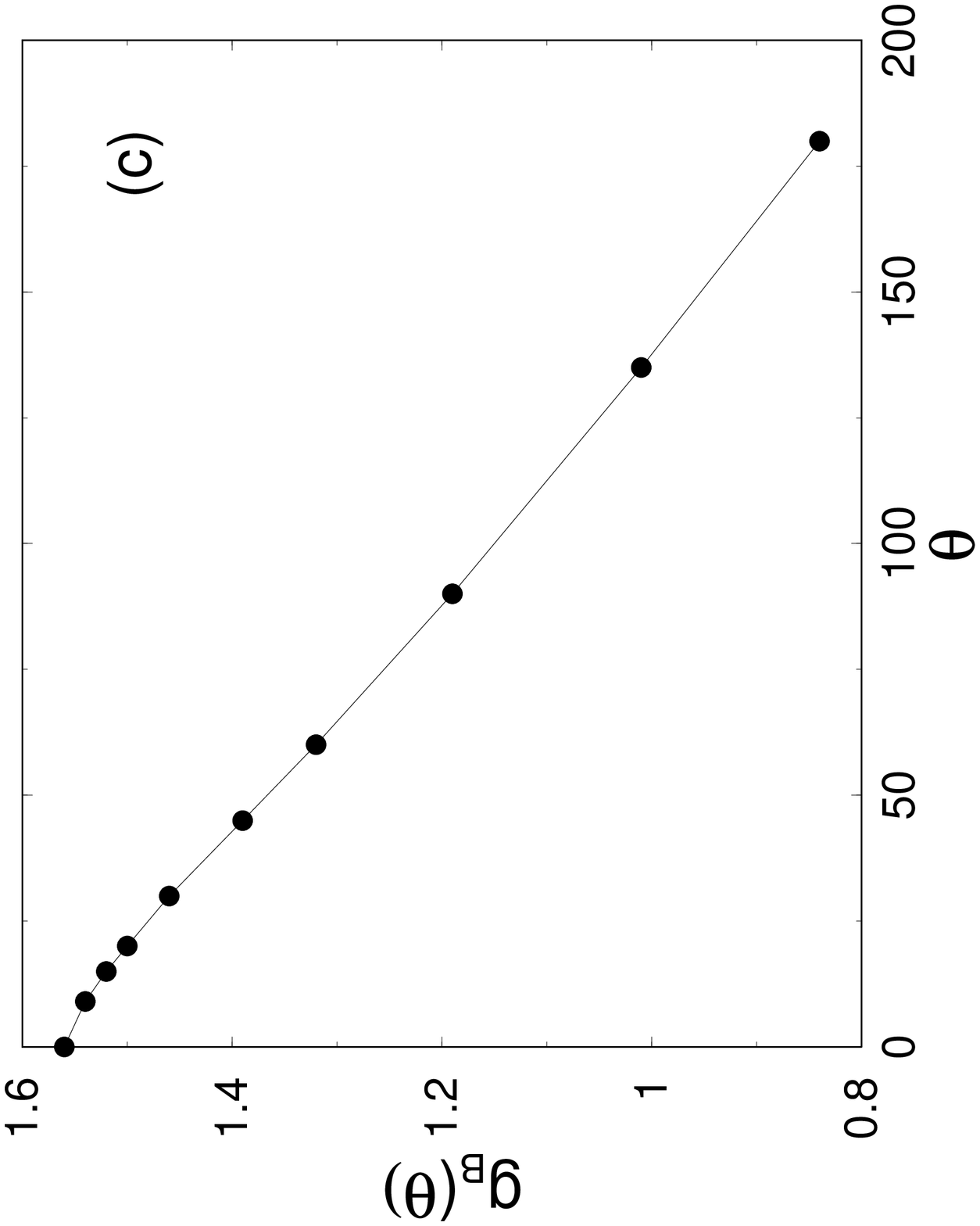}}
}

\caption{(a) Backbone mass distribution $P(m_B|\theta)$ for non-parallel
wells with $r=8$, $w=64$, and several values of $\theta$. The cutoff of
cluster growth is at a cluster mass of $2^{18}$. (b) Backbone mass
distribution $P(m_B|\theta)$ for the case of nonparallel wells with
$r=1$, $w=64$ for various $\theta$. The cutoff of cluster growth is at a
cluster mass of $2^{20}$. (c) Power law exponent $g_B(\theta)$, defined
in Eq.~(\protect\ref{e10x}), for the corresponding backbone mass
distribution presented in (b).}
\label{bbr8thetaX}
\end{figure}

\begin{figure}

\centerline{
\epsfxsize=7.0cm
\epsfclipon
\rotate[r]{\epsfbox{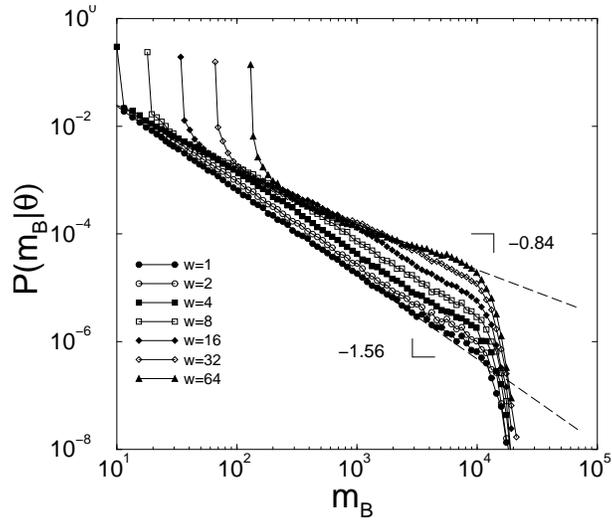}}
}

\caption{Backbone mass distribution $P(m_B|w)$ for the non parallel
wells for fixed $\theta$ ($=180^\circ$) and several values of $w$.  The
larger the value of $w$, the later a crossover occurs from behavior
reflecting a configuration of 2 lines with $\theta=180^\circ$ to a
configuration effectively of 2 points, with power-law-regime exponent
$\tau_B -1$.}
\label{bb1thetaX}
\end{figure}

\end{document}